\documentclass[twocolumn]{aastex}

\usepackage{xcolor}

\fullcollaborationName{The Friends of AASTeX Collaboration}

\begin{document}

\title{The modulation of anomalous and galactic cosmic ray oxygen over successive solar cycle minima}


\author{R.D. Strauss\altaffilmark{1}, R.A. Leske\altaffilmark{2}, J.S. Rankin\altaffilmark{3}}

\altaffiltext{1}{Center for Space Research, North-West University, Potchefstroom, 2522, South Africa} 
\altaffiltext{2}{California Institute of Technology, Mail Code 290-17, Pasadena, CA 91125, USA}
\altaffiltext{3}{Department of Astrophysical Sciences, Princeton University, Princeton, NJ 08540, USA}


\begin{abstract}

Both the recent 2009 and 2020 solar minima were classified as unusually quiet and characterized with unusually high galactic cosmic ray (GCR) levels. However, unlike the trends from previous decades in which anomalous cosmic ray (ACR) and GCR levels strongly agreed, the ACR intensities did not reach such high record-setting levels.This discrepancy between the behaviour of GCRs and ACRs is investigated in this work by simulating the acceleration and transport of GCR and ACR oxygen under different transport conditions. After using recent observations to constrain any remaining free parameters present in the model, we show that less turbulent conditions are characterized by higher GCR fluxes and low ACR fluxes due to less efficient ACR acceleration at the solar wind termination shock. We offer this as an explanation for the ACR/GCR discrepancy observed during 2009 and 2020, when compared to previous solar cycles. 

\end{abstract}


\section{Introduction}

\begin{figure*}[]
\begin{center}
\includegraphics[width=0.99\textwidth]{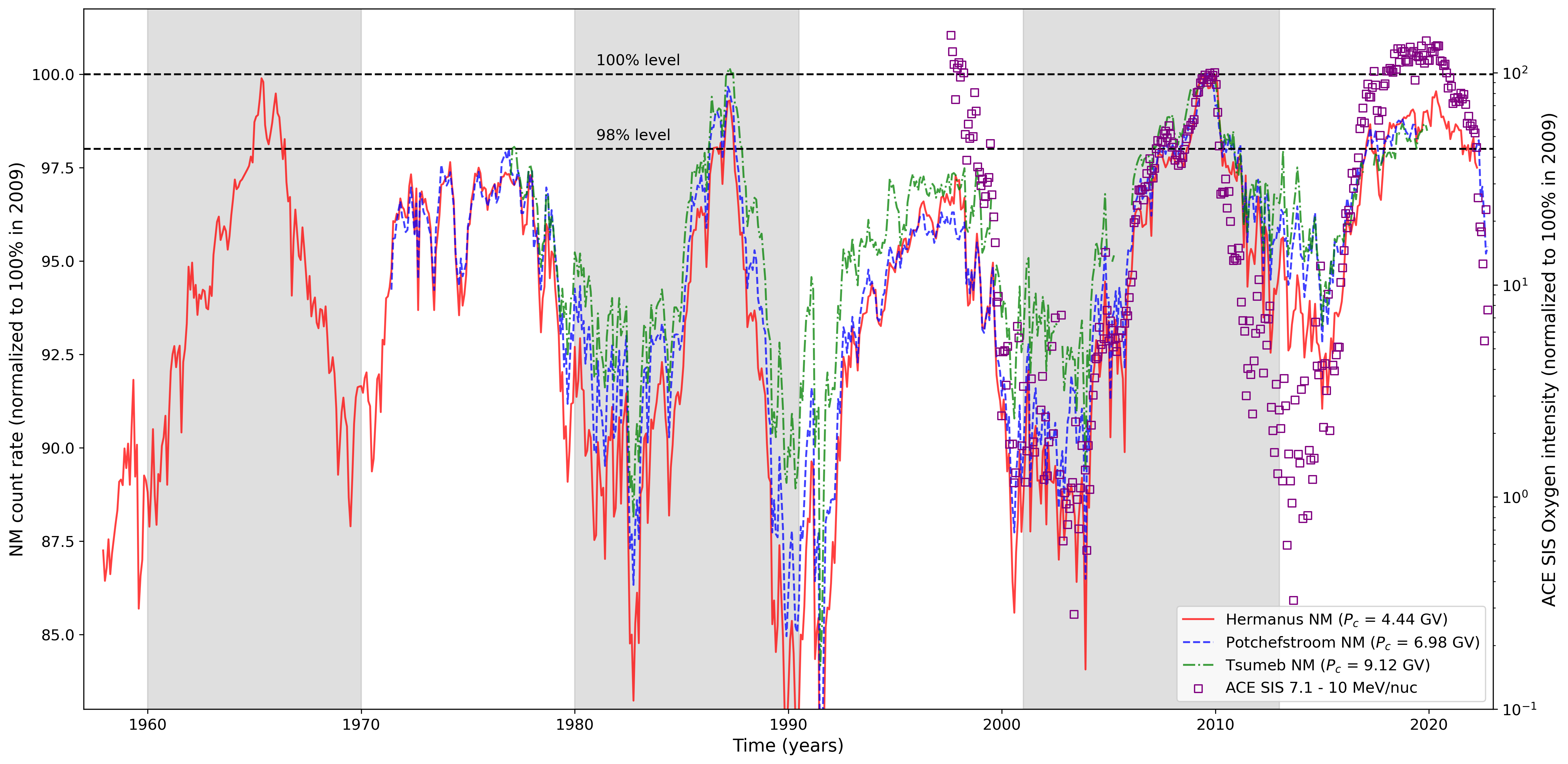}
\caption{Long-term monthly measurements, normalized to 100\% in 2009, from the South African NM network. Different stations are labelled according to their cut-off rigidity, $P_c$. The data is available from \url{http://natural-sciences.nwu.ac.za/neutron-monitor-data}. Also shown are Bartel rotation-averaged quiet time ACE SIS measurements of mainly ACR oxygen at 7.1 - 10 MeV/nuc, again normalized to 100\% in 2009. \label{fig:hermanus}}
\end{center}
\end{figure*}

The 2009 solar minimum was characterized by record-setting galactic cosmic ray (GCR) intensities. \citet{mewaldtetal2010} reported GCR intensities, in the energy range $\sim 70 - 450$ MeV/nuc, to be 20\% -- 26\% greater in late 2009 than in the previous 1997 -- 1998 minimum as well as all the previous solar minima of the space age (1957 -- 1997). Interestingly, the 2009 solar minimum was in a $A<0$ magnetic polarity cycle which is usually associated with lower GCR levels than a similar $A>0$ cycle. This led, amongst others \citet{strausspotgieter2014} and \citet{molotoetal2018}, to postulate that even higher GCR intensities will be observed in the 2020 $A>0$ solar minimum if similarly quiet solar conditions prevailed just due to the drift effect. This was confirmed by \citet{fuetal2021}, showing that GCRs intensities in 2020 were $\sim 6$\% higher than in 2009, and $\sim 25$\% higher than the previous $A>0$ drift cycle in 1997 at energies of $\sim 50 - 500$ MeV/nuc. Record-setting intensities were also reported from ground based neutron monitors (NMs) in 2009 which are sensitive to mostly protons with energies $> 500$ MeV \citep[e.g.][]{moraalstoker2010}. This trend continued into 2020, as illustrated in Fig. \ref{fig:hermanus}. Here we show long-term NM observations, normalized to 100\% in 2009, by the South African NM network, extending back to 1957. In this dataset, the 2009 intensities showed an enhancement of $\sim 1$\% over that of the 1987 $A<0$ cycle, while being only slightly higher than the 1965 $A<0$ cycle. The recent 2020 intensities, in a $A>0$ drift cycle, is however $\sim 2$\% higher than previous $A>0$ maxmima in 1997 and 1975. Note that, because of drift effects, high-energy GCR observations at NM energies are lower in a $A>0$ cycle than a $A<0$ cycle in contrast to low energy space-based measurements, leading to well-known cross-over of the energy spectra \citep[e.g.][]{reineckepotgieter1994}.

\begin{figure*}
\begin{center}
\includegraphics[width=0.99\textwidth]{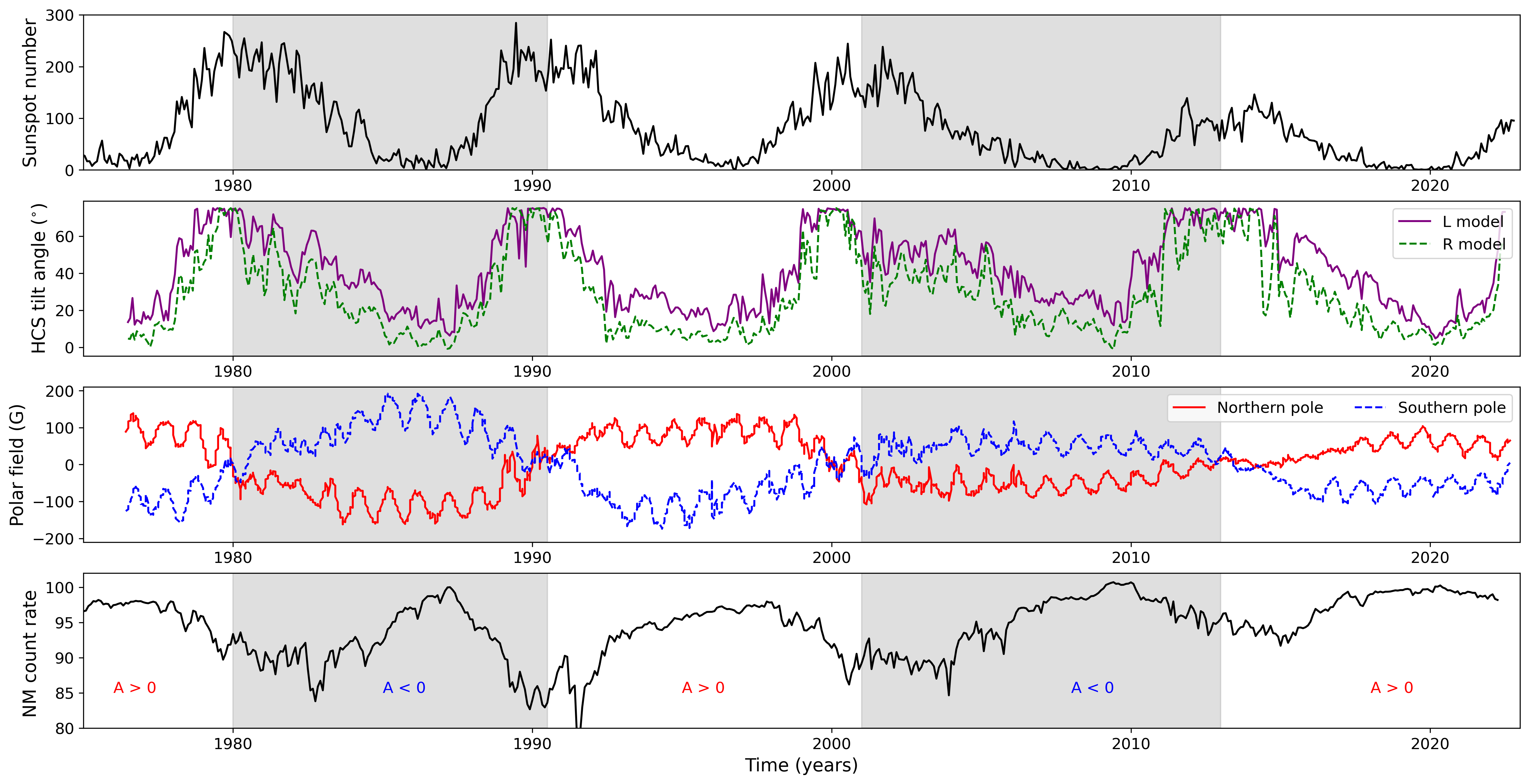}
\caption{The panels show, from top to bottom, the monthly sunspot number (from \url{http://sidc.oma.be/}), the inferred HCS tilt angle for two different models (from \url{http://wso.stanford.edu/Tilts.html}), the magnitude and polarity of the solar polar magnetic field (from \url{http://wso.stanford.edu/Polar.html}) and the Hermanus NM measurements from Fig. \ref{fig:hermanus}. \label{fig:quantities}}
\end{center}
\end{figure*}

Possible reasons for the very high GCR levels in the recent solar cycles have been debated over the past decade. Ultimately the high GCR levels are due to extra-ordinary low levels of modulation caused by the very quiet Sun in both 2009 and 2020. The top panel of Fig. \ref{fig:quantities} shows, as a solar proxy, the sunspot number which exhibits a downward trend over recent cycles with a long and prolonged minimum in both 2009 and 2020. However, the main process influencing GCR modulation is still debated, but it is probably driven by a combination of factors, including a {weaker and} less turbulent magnetic field \citep[][]{smithbalogh2008}, and a slowly declining heliospheric current sheet (HCS) tilt angle \citep[][]{wangetal2009}.

Generally, the intensities of anomalous cosmic rays (ACRs) are observed to track their GCR counterparts rather well. At least, this was the situation for most of the Space Age. A striking feature of the 2009 solar minimum is that, although GCRs reach record setting intensities, ACR intensities only reached ~90\% of their 1997 levels \citep[][]{leskeetal2009,mcdonaldetal2010,leskeetal2013}. This trend continued in 2020 \citep[]{fuetal2021} and also illustrated in Fig. \ref{fig:hermanus}. A natural interpretation for both the record setting GCR intensities, and the fact that ACRs were lower in recent solar cycles, was presented by \citet{moraalstoker2010}: These authors argue that the prevailing quiet solar conditions in 2009 (and therefore also in 2020) lead to less GCR modulation due to lower levels of turbulence that increase the diffusion coefficients (and therefore the mean-free-paths) due to less efficient particle scattering. For ACRs, however, a larger diffusion coefficient leads to less efficient particle acceleration at the solar wind termination shock. While ACRs are therefore also less modulated in 2009 and 2020, their source intensities may also be less, creating an intricate interplay between acceleration efficiency and modulation. This idea has not been robustly tested before and we aim to do so in this current work by simulating the modulation of ACR and GCR oxygen, simultaneously in a numerical modulation model, for different levels of solar modulation.

\section{The acceleration and transport model}

For the simulations presented here, we use the 2D acceleration and transport model of \citet{straussetal2010a, straussetal2010b, straussetal2011} which solves the isotropic \citet{parker1965} transport equation in the meridional plane of the heliosphere. This model is based on earlier work by \citet{lerouxetal1996} and \citet{langneretal2004} {where the Parker transport equation is integrated numerically by a finite difference numerical scheme discussed in detail by \citet{lagnerThesis}}. Diffusive shock acceleration of a pick-up ion source at the solar wind termination shock (TS; assumed to be located at 90 AU) is included, with a latitude independent compression ratio of $s=2.2$. Incompressible plasma flow is assumed in the heliosheath, leading to no adiabatic energy losses/gains in this region, while second-order Fermi acceleration (i.e. momentum diffusion) is also neglected. A solar wind speed of 400 km/s is assumed at Earth in the equatorial plane, increasing to 800 km/s at the polar regions. A \citet{parker1958} heliospheric magnetic field (HMF) is used in the equatorial regions, but modified in the polar regions according to \citet{jokipiikota1989}. {This field is assumed to also be frozen into the slowing heliosheath plasma}. Gradient, curvature, and HCS drifts are included and we use a {constant} HCS tilt angle of $10^{\circ}$ for all simulations {as only solar minimum periods are considered}.

The transport (diffusion and drift) coefficients are discussed in Sec. \ref{Sec:diff_coeff}. It is important to note that the only differences between the GCR and ACR simulations presented here are the charge of the particle population under consideration (fully ionized GCRs and single-ionized ACRs) and the particle source (for ACRs, a pick-up ion source function is specified at the TS, while, for GCRs, a local interstellar spectrum (LIS; see Sec. \ref{Sec:LIS}) is specified at the heliopause (HP), assumed to be located at 120 AU). 

\subsection{Transport parameters}
\label{Sec:diff_coeff}

\begin{figure*}
\begin{center}
\includegraphics[width=0.99\textwidth]{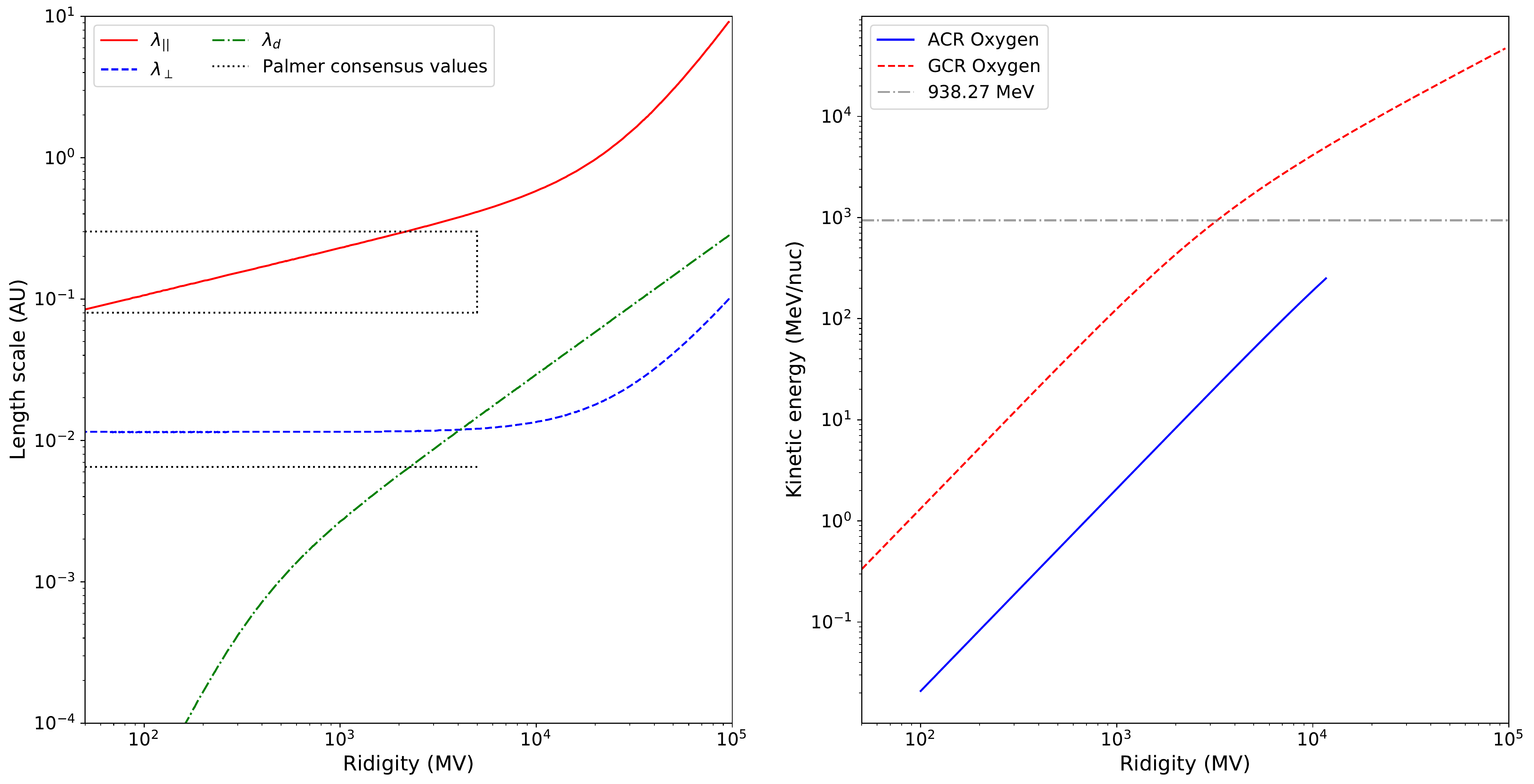}
\caption{The left panel shows the implemented parallel and perpendicular mean-free-paths, along with the drift scale, and so-called consensus values of these quantities for protons. The right panel shows the relationship between rigidity and kinetic energy for the ACR and GCR oxygen components. The horizontal dashed line is the proton rest mass energy. \label{fig:coefficients}}
\end{center}
\end{figure*}

The transport coefficients used in this study are shown, as a function of rigidity, in the left panel of Fig. \ref{fig:coefficients} and briefly discussed below. The parallel and perpendicular coefficients are also compared to the so-called Palmer consensus values \citep[as taken from][]{Bieberetal1994} in order to validate the assumptions used in this study. Note that the diffusion coefficient, $\kappa$, is related to the corresponding mean-free-path, $\lambda$, as $\kappa = v/3 \lambda$ with $v$ particle speed. The right panel of the figure relates the kinetic energy to the rigidity for the different oxygen species considered here. Note that, due to the different charge-to-mass ratio, this relationship is different for the different species.

For the parallel mean-free-path we use the result of \citet{burgeretal2008} and \citet{engelbrechtburger2010}, based on the calculations of \citet{teufelschlickeiser2003},

\begin{equation}
\label{Eq:lambda_parallel}
    \lambda_{||} = \zeta \frac{3}{\pi} \frac{B^2}{\delta B_{\text{slab}}^2} \frac{k_{\min} r_L^2}{\nu - 1} \left[ \frac{1}{4}  + \frac{2 \left( k_{\min} r_L \right)^{-\nu} }{(2 - \nu)(4 - \nu)}  \right].
\end{equation}

In this expression, $B$ is the HMF magnitude, $\delta B^2$ the variance associated with the slab component of the turbulent magnetic field, $r_L$ is the maximal Larmor radius, $k_{\min}$ the wavenumber associated with the transition from the energy to the inertial range of the slab power spectrum, and $\nu = 5/3$ the spectral index of the inertial (Kolmogorov) turbulence range. Following \citet{burgeretal2008} we use $k_{\min}=32 \text{ AU}^{-1} \sqrt{r}$, where $r$ is radial distance. {The value of $k_{\min}$ decreases until the TS is reached and kept constant in the heliospheath.} Similarly we assume that $\delta B_{\text{slab}}^2 = 13.2 \text{ nT} r^{-2.5}$ inside the TS but $\delta B_{\text{slab}}^2 \propto B^2$ further into the heliosheath. The free parameter $\zeta$ is introduced to account in an {\it ad hoc} fashion for time-dependent changes in $\lambda_{||}$ \citep[see e.g. the discussion by][]{strausspotgieter2010} and is fixed by comparing model results to observations.

For the perpendicular diffusion coefficients, $\kappa_{\perp, r\theta}$ we use the estimate of \citet{burgeretal2000},

\begin{equation}
\label{kappa_perp}
\kappa_{\perp/r} = \eta \kappa_{||} \left[ \frac{P}{\text{1 GV}} \right]^{-1/3} \text{ and } \kappa_{\perp,\theta} = \eta g(\theta) \kappa_{||} \left[\frac{P}{\text{1 GV}}  \right]^{-1/3}
\end{equation}

Note that the perpendicular diffusion coefficient in the polar region, $\kappa_{\perp,\theta}$ is enhanced with respect to the radial direction, $\kappa_{\perp,r}$, via the function $g(\theta)$, leading to isotropic perpendicular diffusion in the equatorial plane but becoming increasingly anisotropic, so that $g=4$ at the polar regions. This enhancement has been implemented in order to reproduce observed latitudinal gradients \citep[e.g.][]{potgieteretal1977}. The parameter $\eta$ is adjusted in the following sections in order to reproduce 1 AU observations.

The efficiency of drift effects can be characterized by the so-called drift scale, $\lambda_d$, which reduces to the Larmor radius if scattering is negligible \citep[e.g.][]{minnietal2007,burgervisser2010,engelbrechtetal2017,jabusdrift}. If, however, solar wind turbulence is present, $\lambda_d = f_s r_L$, with $f_s \in[0,1]$ referred to as the drift reduction factor. Here we use a parameterized form of this reduction factor from \citet{burgeretal2000} and \citet{langneretal2004},

\begin{equation}
    f_s = \chi \frac{(P/P_0)^2}{(P/P_0)^2 + 1},
\end{equation}

with $P_0 = 1/10$ GV. Again, $\chi$ will be fixed by comparing simulation results with observations.

{In constrast to incompressible turbulence observed in the supersonic solar wind, turbulence in the heliosheath has been observed to be mostly compressible \citep[e.g.][]{burlaganess2009}. Presently, a complete theory of particle scattering (both parallel and perpendicular to the mean field) in compressive turbulence does not exists and, as such, Eqs. \ref{Eq:lambda_parallel} and \ref{kappa_perp} are assumed to be valid throughout the heliosphere and also applied in the heliosheath. }

\subsection{The oxygen LIS}
\label{Sec:LIS}

\begin{figure}
\begin{center}
\includegraphics[width=0.49\textwidth]{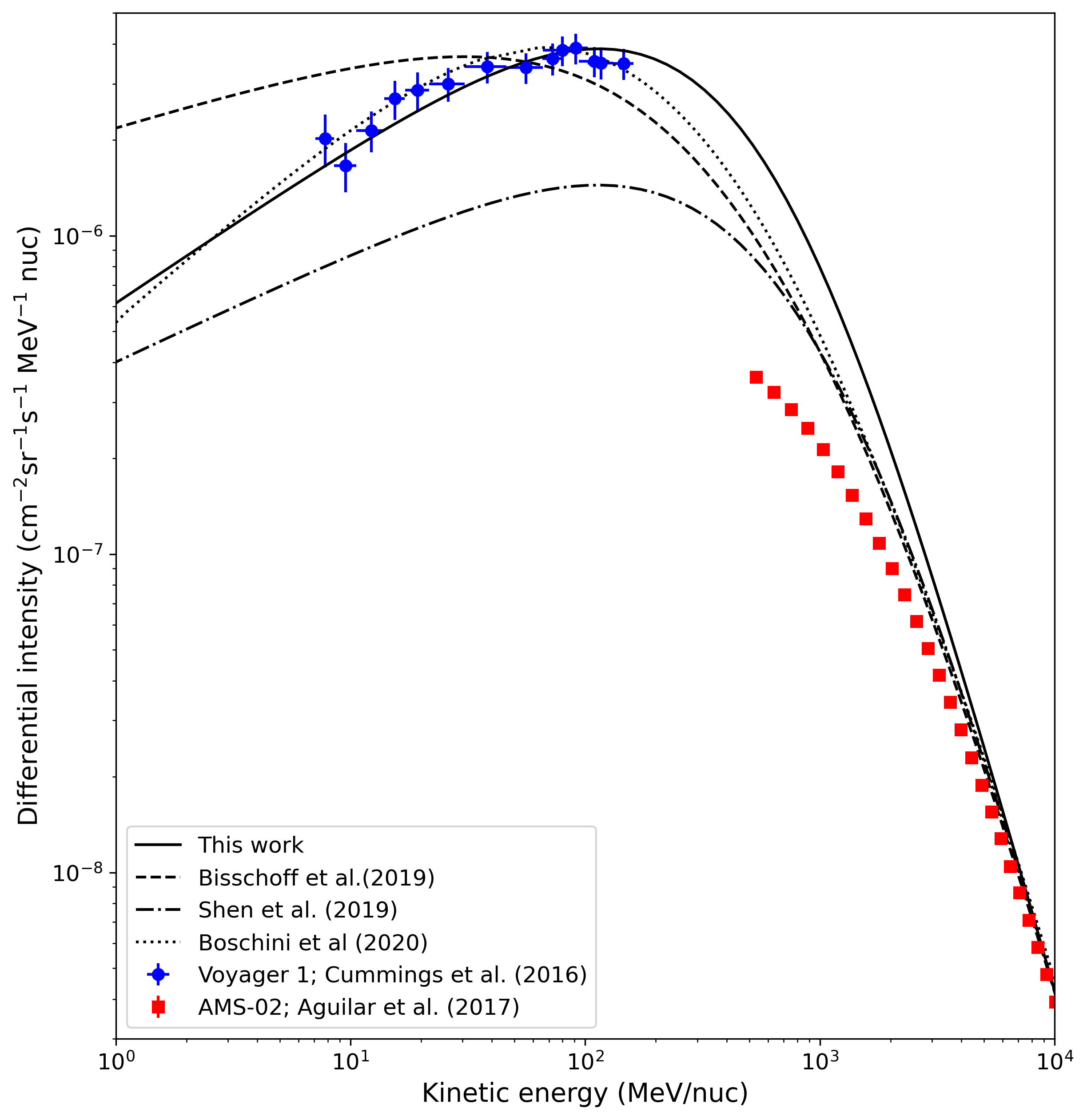}
\caption{A comparison of several contemporary oxygen LIS estimates, as compared to Voyager and AMS-02 measurements. \label{fig:LIS_various}}
\end{center}
\end{figure}

For the GCR oxygen simulations we assume a LIS that is comparable to the available observations, while having a rather simplistic mathematical form,

\begin{equation}
    j_{LIS} = j_0 \beta^{\gamma} \left( \frac{E}{1 \text{ GeV/nuc}} \right)^{- \delta} 
\end{equation}

where $\beta$ is the ratio of particle speed to the speed of light, $E$ is the kinetic energy per nucleon, $\gamma = 6.25$, $\delta = 2.625$, and $j_0 = 6250$ cm$^{-2}$s$^{-1}$sr$^{-1}$MeV$^{-1}$nuc.  In Fig. \ref{fig:LIS_various} this LIS is compared to other estimates by \citet{Boschinietal2018}, \citet{Bisschoffetal2019}, and \citet{shetetal2019}, where the latter should rather be interpreted as a source function, specified in their simulations at 80 AU. Also shown in the figure, is high-energy AMS-O2 data \citep[][]{Aguilaretal2017}, taken between 2011 -- 2016, and low-energy Voyager measurements beyond the HP \citep[][]{Cummingsetal2016} used to constrain the LIS.

\begin{figure*}
\begin{center}
\includegraphics[width=0.99\textwidth]{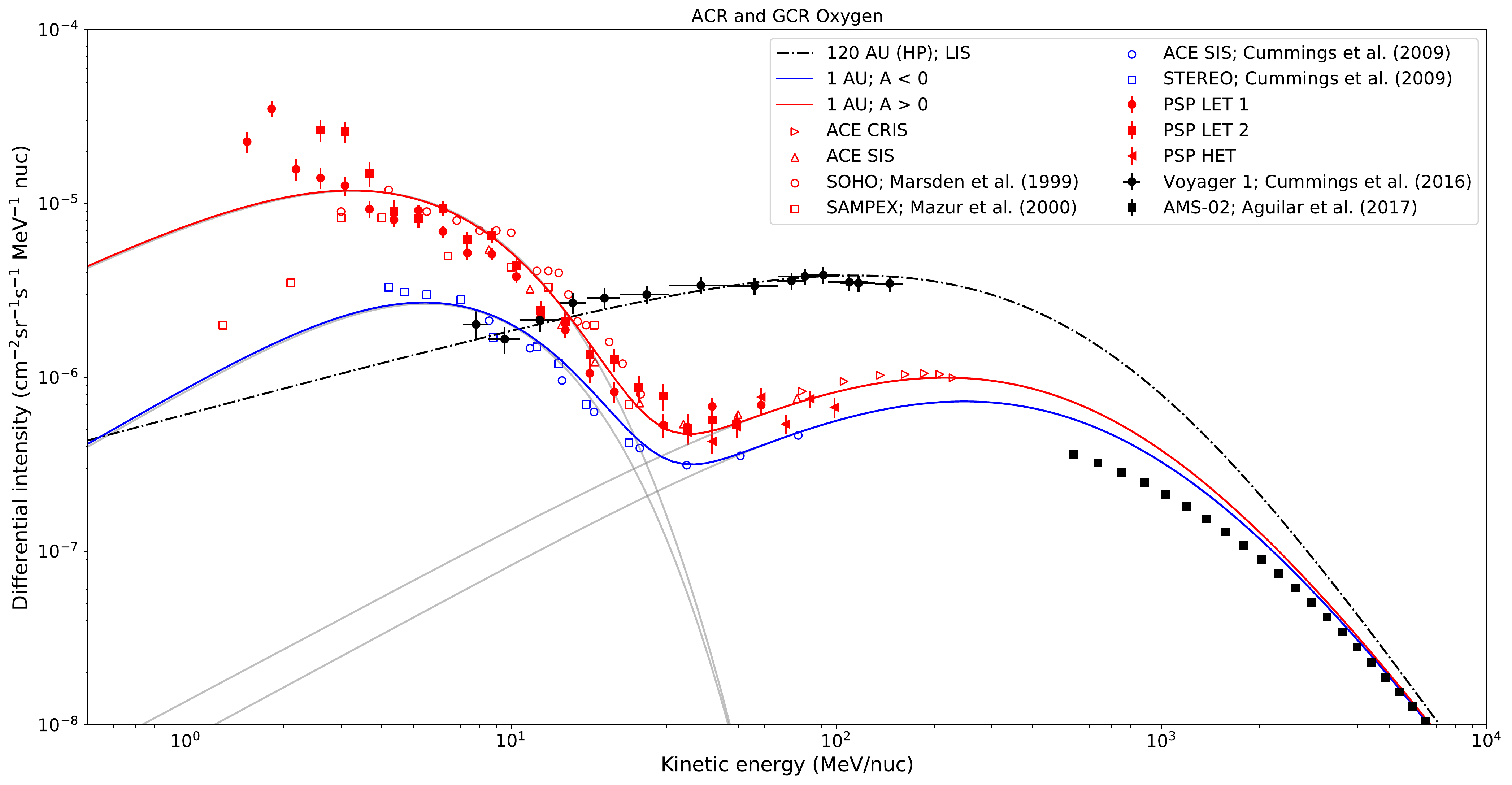}
\caption{The red symbols show the $A>0$ measurements used to constrain the simulations, while $A<0$ observations are shown by blue symbols. Similarly, simulations for the $A>0$ drift cycles are shown in red and $A<0$ simulations in blue. The data and simulations are for Oxygen.  \label{fig:spectra}}
\end{center}
\end{figure*}

\section{Reproducing anomalous and galactic Oxygen spectra}
\label{Sec:model_tuning}

We start by using GCR oxygen energy spectra at Earth to constrain the remaining free parameters in the transport model. To do so, we use the highest energy channels of the ACE CRIS instrument \citep[][]{stoneetal1998} from 2008 and 2020, along with recent PSP ISOIS \citep[][]{mccomas2016} data from the EPI-Hi sensor in 2020 \citep[][]{rankinetal2021b}. The model is then run for different values of the transport parameters until a reasonably good fit is obtained for both polarity cycles. We find optimal agreement with $\zeta = 1.08$, $\eta = 0.05$, and $\chi = 0.6$. Here, $\zeta > 1$ suggests that both the 2009 and 2020 solar minimum were less active than previous cycles and is generally consistent with lower turbulence levels. A value of $\eta = 0.05$ for the ratio of perpendicular to parallel diffusion is about a factor of 2 higher than previous estimates \citep[e.g.][]{ferreirapotgieter2004} but still reasonable considering how much this ratio depends on the level of turbulence \cite[see e.g.][]{giacalonejokipii1999} which could be very different for the recent quiet solar cycles. A drift reduction factor of $\chi = 0.6$ is in line with previous estimates of this quantity from e.g. \citet{langneretal2004} who found a value of $0.55$ for this coefficient during solar minimum conditions.

With all free parameters now determined, we run the model for ACR oxygen in both polarity cycles. As the exact level of the ACR oxygen seed population at the TS is not known, the simulations must, unfortunately, be normalized to 1 AU levels. Note that the same normalization factor is used in simulations in both polarity cycles. For the normalization we use $A>0$ measurements in 2020 from PSP ISOIS (EPI-Hi) \citep[][]{mccomas2016} and ACE SIS \citep[][]{stoneetal1998b}. However, as shown by \citet{rankinetal2021b}, ACR oxygen spectra for different solar cycles compare extremely well, allowing us to also use data from the 1990s, including SOHO measurements from \citet{marsdenetal1999} and SAMPEX measurements from \citet{mazuretal2000}. For $A<0$ measurements, we use ACE and STEREO measurements \citep[][]{cummingsetal2009}.

Our best fit scenario, for both ACR and GCR oxygen, is now shown in Fig. \ref{fig:spectra}. Here, $A>0$ simulations and measurements are shown in red, while $A<0$ simulations and measurements are shown in blue. The individual ACR and GCR simulations are added in grey. The simulations compare very well with both the ACR and GCR measurement for both polarity cycles. {For the GCR Oxygen spectra presented here we do not observe the cross-over for different drift cycles alluded to earlier. While this can potentially occur at higher energies than considered here, its absence may point to an inaccurate form of our assumed transport coefficients at the highest energies, or it could be that such a cross-over does not exist for Oxygen. However, such high energy discrepancies will not influence the rest of the results presented here.}

\section{The effect of changing transport conditions}

\begin{figure*}
\begin{center}
\includegraphics[width=0.99\textwidth]{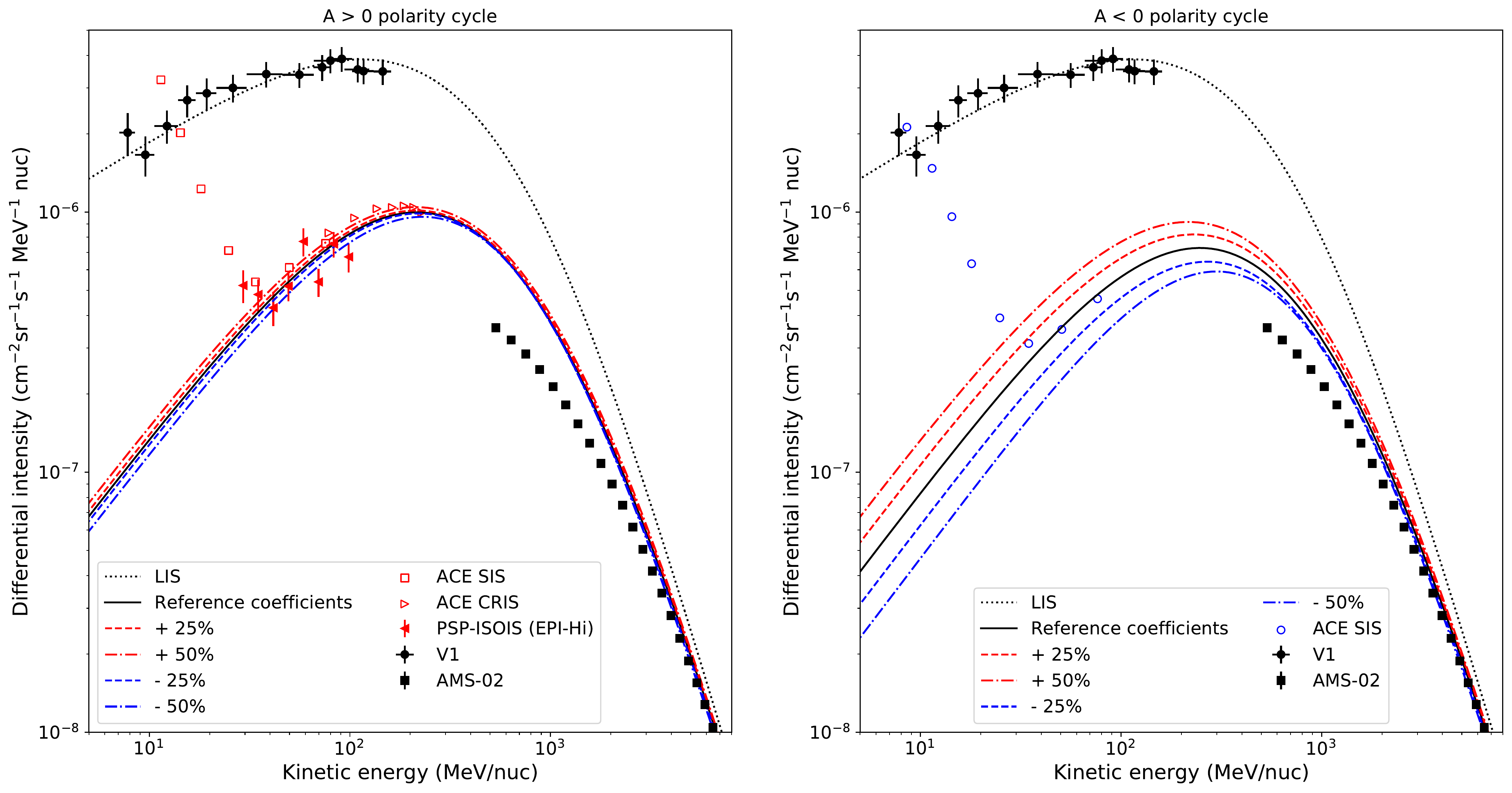}
\includegraphics[width=0.99\textwidth]{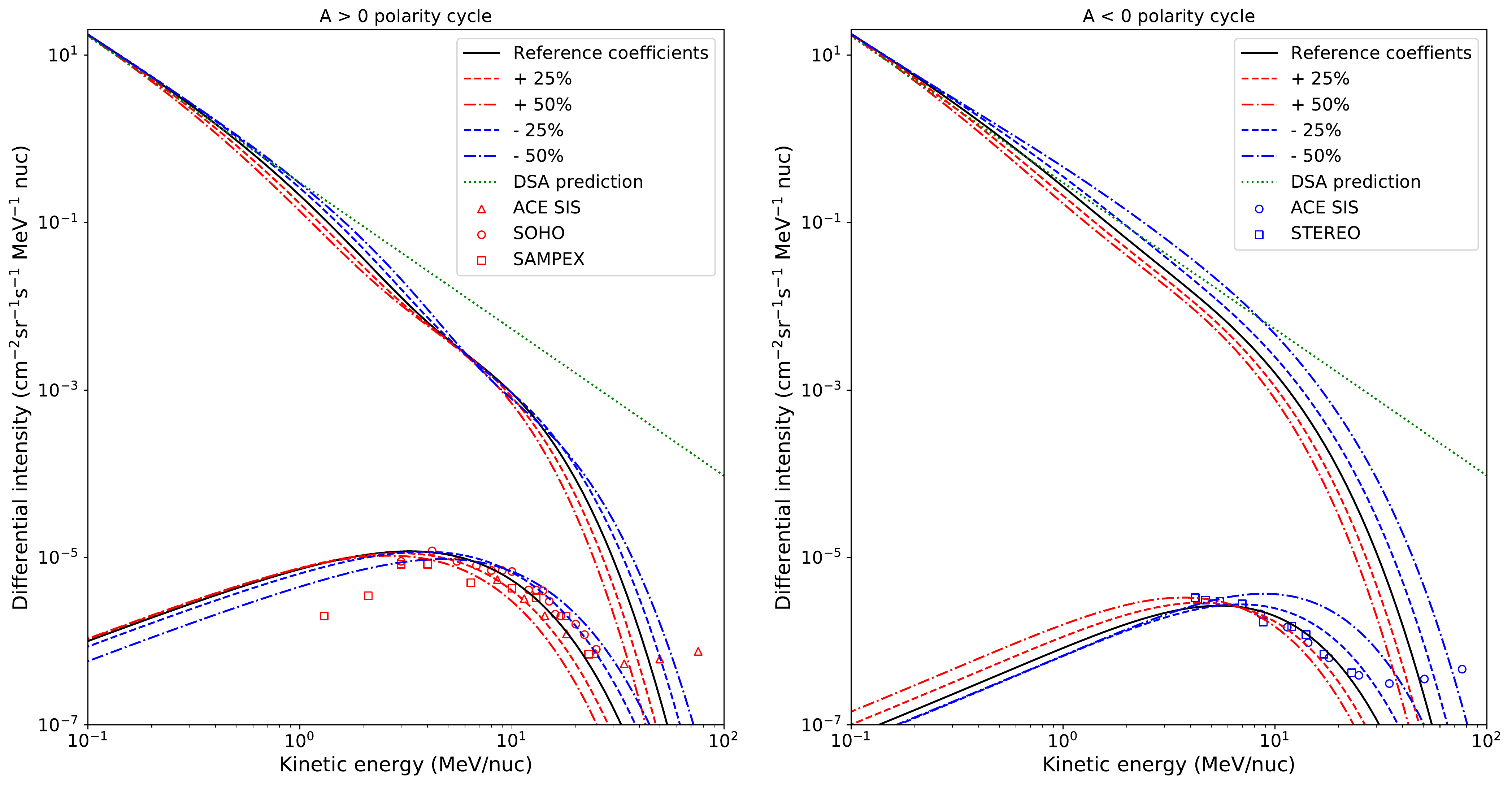}
\caption{The response of simulated GCR (top panels) and ACR (bottom panels) energy spectra to changing diffusion coefficients for $A>0$ (left panels) and $A<0$ (right panels) drift cycles. For the ACR simulations, the power-law prediction from DSA theory (green dotted line) is also included. \label{fig:changing_spectra}}
\end{center}
\end{figure*}

To show the effect of changing diffusion conditions on the resulting ACR and GCR spectra at 1 AU, we vary the value of $\zeta$ by $\pm 50\%$ and show the resulting spectra in Fig. \ref{fig:changing_spectra}. Here, the top panels correspond to GCR simulations and the bottom panels to ACR simulations. Left panels are for the $A>0$ polarity cycle, while right panels are for the $A<0$ polarity cycle. The result for GCR oxygen is rather straightforward; a larger diffusion coefficient, which corresponds to less turbulent conditions, leads to higher GCR levels with slight differences between different drift cycles. Most notably, the GCR intensity is much more sensitive to changes in the assumed turbulence conditions during $A<0$ cycles than during $A>0$ cycles.

For ACR oxygen, however, an interesting interplay is obtained between the ACR source spectrum at the TS and the resulting modulated spectrum at Earth. At the TS, a smaller diffusion coefficient, which corresponds to more turbulent conditions, lead to an accelerated spectrum extending to higher energies. This is a well-known result for diffusive shock acceleration \citep[DSA, see e.g.][]{SteenbergMoraal1999,moraalstoker2010,prinbslooetal2019}; a smaller diffusion coefficient leads to better particle confinement at/near the shock, allowing the DSA process to continue to higher energies before an exponential cut-off is observed. In general, therefore, more turbulence (i.e. more particle scattering) leads to more efficient ACR acceleration at the TS. However, this resulting harder energy spectrum will now be modulated more efficiently by the additional turbulence as the ACR particles propagate towards Earth. In the bottom panels of Fig. \ref{fig:changing_spectra} this interplay between more efficient acceleration {\it vs.} more effective modulation is illustrated for both drift cycles. {More turbulent conditions (i.e. a smaller diffusion coefficient) leads to higher ACR intensities at the highest ACR energies ($\gtrsim 5$ MeV/nuc) at both the TS and at Earth. For low energy ACRs at Earth, however, a smaller diffusion coefficient leads to lower ACR intensities.}

\begin{figure*}
\begin{center}
\includegraphics[width=0.99\textwidth]{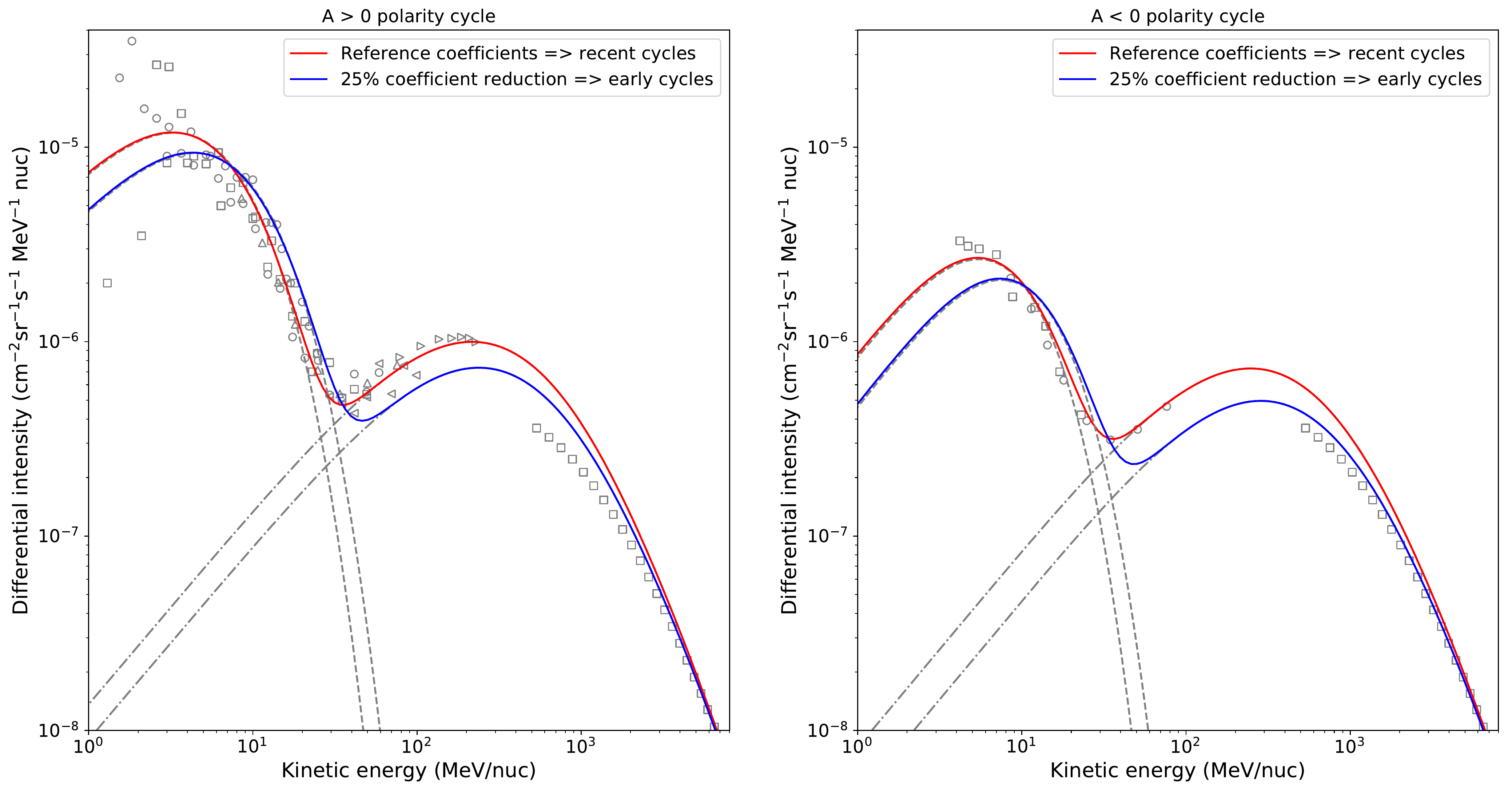}
\includegraphics[width=0.99\textwidth]{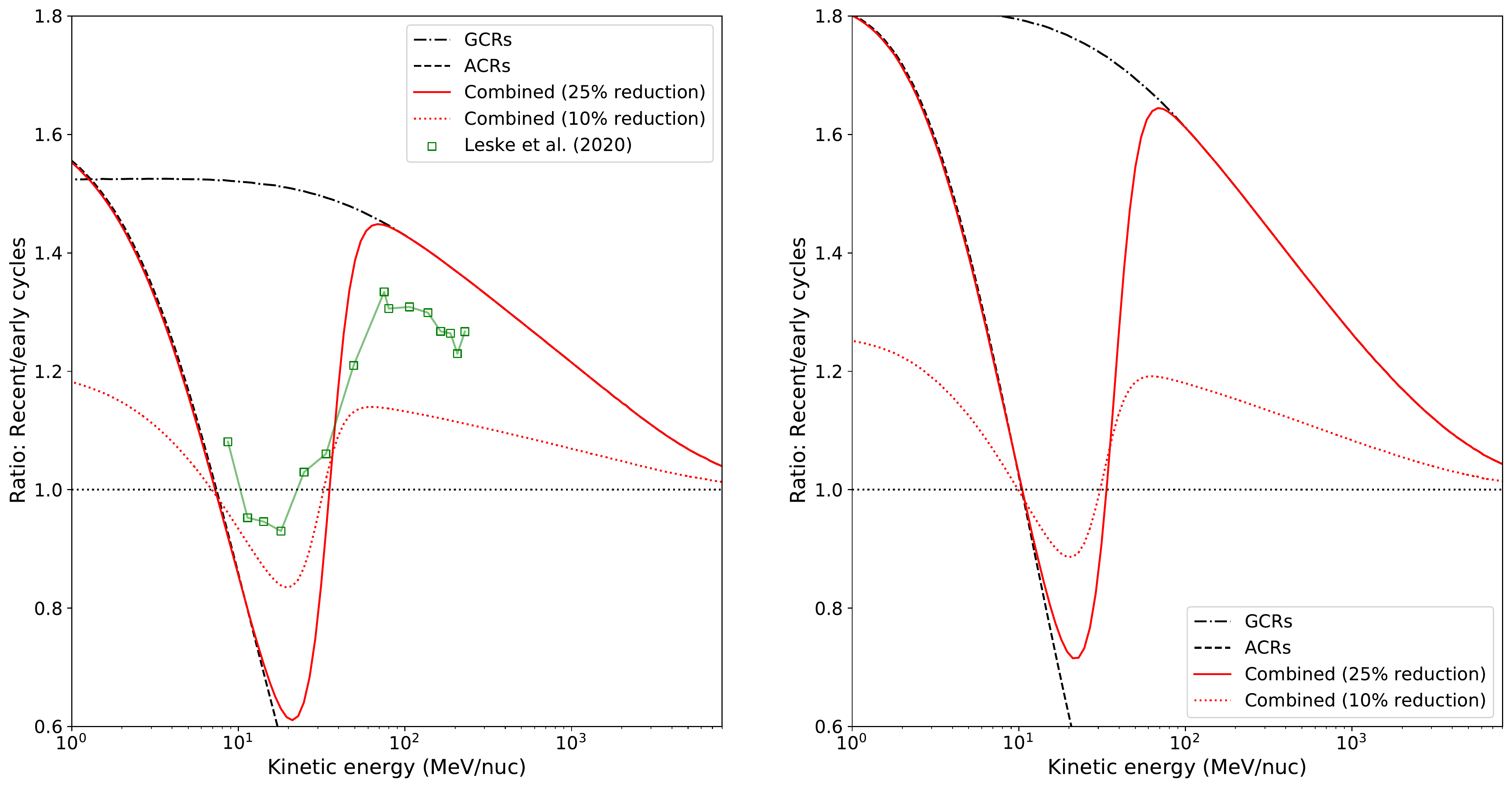}
\caption{The top panels show ACR and GCR simulations for $A>0$ (left panels) and $A<0$ (right panels) drift cycles for so-called {\it recent} (red) and {\it early} (blue) solar cycles. The bottom panels show the ratio of the energy spectra for these two scenarios. \label{fig:spectra_ratio}}
\end{center}
\end{figure*}

We can now use the preceding simulations to address the observations by \citet{leskeetal2013}, who found higher GCR, but lower ACR, intensities during the recent, quiet, solar minima. The simulations presented in Fig. \ref{fig:spectra} are representative of the quiet solar conditions that were observed in both 2009 and 2020, and we refer to these as results for {\it recent} solar cycles. Next, we proposed that {\it early} solar cycles, i.e. the solar cycles before 2009 were generally more active (i.e. more turbulent) than these recent cycles and assume that the transport coefficients (both the diffusion and drift coefficients) were $25\%$ smaller in early cycles as compared to recent solar cycles. The resulting simulations are shown in the top panel of Fig. \ref{fig:spectra_ratio} for the $A>0$ (left panels) and $A<0$ (left panels) polarity cycles. Results for {\it recent solar cycles} are shown in red, while blue corresponds to that of {\it early} solar cycles with a $25\%$ reduction in all transport coefficients. As expected from the previous simulations, this results in GCR intensities that are higher in recent solar cycles, but ACR intensities for $E \gtrsim 5$ MeV, being higher in earlier cycles. At the lowest energies we again find higher intensities in recent cycles.

This behaviour is best illustrated in the bottom panels of Fig. \ref{fig:spectra_ratio} where we show the ratio of modelled intensities for recent/early cycles. We do so for ACRs (black dashed line) and GCRs (dashed-dotted line) separately, but also for the combined ratio (solid red line). For completeness' sake, we also repeat the calculation assuming only a $10\%$ reduction in the transport coefficients and show this as the red dotted line. {Also shown, as the green curve, is the measured ratio from ACE (ratio of 2018 -- 2020 to 1997 -- 1998), showing a very good qualitative comparison with the model results.}

\section{Summary and conclusions}

We show modelling results where the same set of transport coefficients are used to reproduce combined ACR and GCR measurements, at Earth, for two recent solar minima. As the GCR oxygen LIS has now been directly observed by the Voyager spacecraft \citep[][]{Cummingsetal2016,stoneetal2019}, a comparison between GCR simulations and observations allows us to constrain the remaining free parameters in the modulation model. The model is then used to also simulate ACR spectra, which we find to be consistent with recent observations.

We study the effect of changing diffusion conditions on energy spectra and find that GCR behave as expected with larger diffusion coefficients, associated with more quiet solar conditions, leading to higher GCR fluxes during recent solar cycle minima. For ACRs, this is more complex as the acceleration of ACRs at the TS is also effected by the magnitude of the diffusion coefficient, with more turbulent conditions leading to more efficient ACR acceleration, while the ACR will be more effectively modulated while propagating towards Earth. Our simulations suggest that high energy ACR intensities will therefore be lower during very quiet solar minimum conditions. At the lowest ACR energies, however, ACR intensities are again higher, illustrating the interplay between acceleration and modulation and their opposite depending on turbulence (scattering) conditions.

By assuming that {\it recent} solar cycles are associated with less scattering, we calculate the combined ACR and GCR spectral ratio between {\it recent} and {\it early} solar cycles. Our results are in agreement with observations presented by \citet{leskeetal2013} showing higher GCR and lower ACR fluxes during the recent quiet conditions. This is also in line with the proposed explanation by by \citet{moraalstoker2010} which is now robustly tested with simulations. \\

\acknowledgments

 This work is based on the research supported in part by the National Research Foundation of South Africa (NRF grant numbers 119424, 120345, and 120847). Opinions expressed and conclusions arrived at are those of the authors and are not necessarily to be attributed to the NRF. The responsibility of the contents of this work is with the authors. R.A.L. acknowledges support from NASA grant 80NSSC18K0223. This work was also supported as a part of the Integrated Science Investigation of the Sun on NASA’s Parker Solar Probe mission, under contract NNN06AA01C. Figures prepared with Matplotlib \citep{hunter} and certain calculations done with NumPy \citep{harrisetal2020}. Sunspot number measurements provided courtesy of the Royal Observatory of Belgium, Brussels. HCS tilt angle and solar polar magnetic field measurements provided courtesy of the Wilcox Solar Observatory (WSO). WSO data used in this study was obtained via the web site \url{http://wso.stanford.edu} at 2022:11:04\_18:40:53 PDT courtesy of J.T. Hoeksema. The WSO is currently supported by NASA.


\end{document}